\documentclass{ws-mpla}

\begin{document}

\markboth{X. Qian}
{SSA/DSA in Pion Electroproduction through DIS}

\catchline{}{}{}{}{}

\title{Single/Double-Spin Asymmetry Measurements of Semi-Inclusive
Pion Electroproduction on a Transversely Polarized $^3$He Target through 
Deep Inelastic Scattering}

\author{\footnotesize Xin Qian\footnote{email:xqian@caltech.edu} \\
on behalf of the Jefferson Lab \\
Hall A Collaboration and E06-010 Collaboration}

\address{Kellogg Radiation Laboratory, California Institute of Technology, \\1200 East
California Boulevard, Pasadena, California 91125, U.S.A.\\
}


\maketitle

\pub{Received (Day Month Year)}{Revised (Day Month Year)}

\begin{abstract}

Parton distribution functions, which represent the flavor and spin structure 
of the nucleon, provide invaluable information in illuminating quantum chromodynamics in the confinement 
region. Among various processes that measure such parton distribution
functions, semi-inclusive deep inelastic scattering is regarded as one of the
 golden channels to access transverse momentum dependent parton distribution functions, which provide 
a 3-D view of the nucleon structure in momentum space. 
The Jefferson Lab experiment E06-010 focuses on measuring the target 
single and double spin asymmetries 
in the $\overrightarrow{^{3}\rm He}(e,e'\pi^{+,-})X$ reaction with 
a transversely polarized $^3\rm He$ target in Hall A with a 5.89 GeV electron beam.
A leading pion and the scattered electron are detected in coincidence by the left High-Resolution 
Spectrometer at $16^\circ$ and the BigBite spectrometer at $30^\circ$ beam right, respectively. The 
kinematic coverage concentrates in the valence quark region, x $\sim$0.1-0.4, at $Q^2$ $\sim$ 1-3 GeV$^2$. 
The Collins and Sivers asymmetries of $^3\rm He$ and neutron are extracted. In this review, 
an overview of the experiment and the final results are presented. Furthermore, an upcoming 12-GeV 
program with a large acceptance solenoidal device and the future possibilities at an 
electron-ion collider are discussed. 

\end{abstract}

\ccode{PACS Nos.: include PACS Nos.}

\section{Introduction}
One of the important tasks of nuclear physics has been and still is to understand
the internal structure of nucleons in terms of quarks and gluons, which 
are the fundamental degrees of freedom of Quantum Chromodynamics (QCD).
QCD, the widely accepted theory of the strong interaction, 
has been well-tested through perturbative calculations at high energies 
where quarks and gluons behave like free particles\cite{asy1,asy2}. 
However, perturbative calculations break down at low energies, 
where colored quarks and gluons are bounded inside colorless hadron, 
since the strong interaction coupling constant increases with the decrement 
of the energy scale. Therefore, the internal structure of nucleons 
still remains illusive. Although a full description of nucleon structure
is not calculable yet, experimental measurements of the universal parton 
distribution functions (PDFs)\cite{hera1} have provided unique and quantitative  information
about the partonic dynamics. PDFs, which are probability densities of finding 
a parton inside a hadron, bridge nucleons and their partonic 
structure together with the fragmentation functions (FFs), which are 
probability densities that a parton hadronize into a hadron. 
In particular, the energy evolution of PDFs has been one of the best tests of QCD.

In the leading twist, there are three quark distribution functions after integrating over the 
quark transverse momentum: the unpolarized PDF $f_1$, the longitudinal polarized PDF $g_1$, and the 
quark transversity PDF $h_1$. Through several decades of experimental and theoretical efforts\cite{hera1}, 
the unpolarized PDF $f_1$ has been extracted with an excellent precision over a large
range of $x$ and $Q^2$ from DIS, Drell-Yan, and other processes. With advances in experimental 
techniques of polarizing leptons and nucleons\cite{spin}, the longitudinal polarized PDF $g_1$ 
has also been extracted with a reasonable precision over a range of $x$ and $Q^2$. The least known
among these three PDFs is the chirally odd transversity PDF $h_1$, which was initially discussed  
by Hidaka, Monsay, and Sivers\cite{trans1} in 1978, by Ralston and Soper\cite{trans2} 
in 1979, and later by Jaffe and Ji\cite{trans3} in early 1990s. The lowest moment of $h_1$ is 
called the ``tensor charge'', which is a fundamental property of the nucleon, and has been 
calculated from lattice QCD\cite{tc_lqcd} and various 
models\cite{tc_mod1,tc_mod2,tc_mod3,tc_mod4,tc_mod5,tc_mod6}. Transversity is further 
constrained by Soffer's inequality\cite{soffer}, $\left|h_{1}^q\right|\le \frac{1}{2}\left(f_1^q + g_1^q\right)$,
which holds under the next-to-leading-order QCD evolution\cite{Kumano:1997qp,Hayashigaki:1997dn,werner}. 
However, a possible violation of Soffer's bound has been suggested\cite{doubt}.

Compared to the collinear PDF $f_1$, $g_1$, and $h_1$, their corresponding TMDs 
depend not only on the longitudinal 
momentum fraction $x$, but also on the parton transverse momentum $k_T$. TMDs provide a full 3-D 
view of the nucleon structure in momentum space.  For example, an intuitive
interpretation of the unpolarized TMD $f_1$ is that it represents the probability of finding a 
parton inside a nucleon with a longitudinal momentum fraction $x$ and a transverse momentum 
$k_T$. Beside TMDs $f_1$, $g_1$, and $h_1$, there are five
more transverse momentum dependent distribution functions (TMDs) in the leading twist\cite{tmd1,tmd2,tmd3}. 
They are the Sivers function $f^{\perp}_{1T}$, the Boer Mulders function $h^{\perp}_1$, 
the transversal helicity function $g_{1T}$, the longitudinal transversity function $h^{\perp}_{1L}$,
and the pretzelosity functions $h_{1T}^\perp$. All eight TMDs and their isospin structure have been studied in the large-$N_c$ QCD~\cite{largenc}. 
Furthermore, these additional five TMDs require interferences 
between wave function components with different amounts of orbital angular momentum 
(OAM)\cite{brodsky,oam}, and thus require non-zero OAM. The Sivers function $f_{1T}^\perp$ 
(the Boer Mulders function $h_{1}^\perp$) provides information
about the correlation between the quark OAM and the nucleon (the quark) spin.
In comparison, the longitudinal function $g_{1L}$ and the transversity function $h_{1T}$ 
describe correlations between the quark spin and the nucleon spin. Furthermore, 
the Sivers function $f_{1T}^\perp$ and the Boer Mulders function $h_{1}^\perp$
are T-odd functions, which rely on the final state interactions (FSI) 
experienced by the active quark in semi-inclusive DIS (SIDIS) process as both 
functions vanish without FSI. On the other hand, the transversal helicity function 
$g_{1T}$, the longitudinal transversity function $h_{1L}^\perp$ and the  
pretzelosity functions $h_{1T}^\perp$ are T-even 
and as a result do not require FSI to be nonzero. The wealth of information from all these functions 
could thus provide invaluable information about the quark orbital
angular momentum. 

\section{Single and Double Spin Asymmetries in SIDIS}

\begin{figure}[]
\centering
\includegraphics[width=90mm]{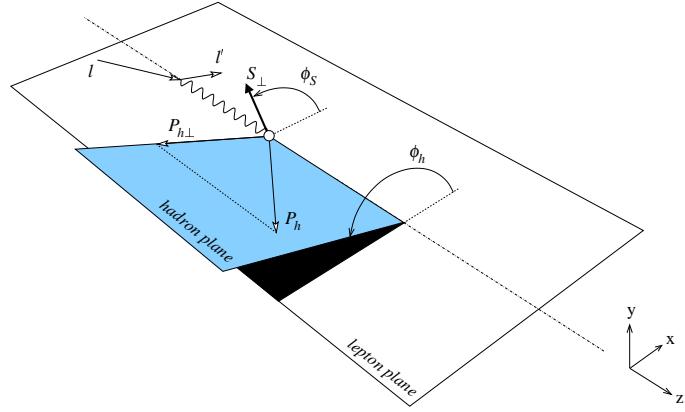}
\caption{\label{fig:angle} Definitions of the azimuthal angle $\phi_h$ and $\phi_S$,
and the hadron transverse momentum for SIDIS process in the nucleon-at-rest 
frame following the Trento convention\protect\cite{trento}. Here, $l$, $l'$, $P_h$, and 
$S_{\perp}$ represent the momenta of the initial lepton, the scattered lepton, 
the leading hadron, and the spin vector of the initial nucleon, respectively.}
\end{figure}

Compared to inclusive DIS process, in which only the scattered lepton is detected, 
SIDIS process $N(l,l'h)X$ also requires detection of one of the leading hadrons 
fragmented from the struck quark. By using the leading hadron 
to tag the flavor, the spin, and the transverse momentum information of the struck quark,
SIDIS process provides unique sensitivities to TMDs. 
Fig.~\ref{fig:angle} shows the kinematics of SIDIS process following the Trento 
convention\cite{trento}. The SIDIS reaction is regarded as one of the golden 
channels to access TMDs, since it can access all eight leading twist TMDs with 
different combinations of beam and target polarizations\cite{tmd3}. 

For example, the angular dependence of the single spin asymmetry (SSA) 
$A^{SSA}$ in the scattering of an unpolarized lepton beam off a transversely polarized 
target is:
\begin{eqnarray}\label{eq:asy}
A^{SSA}(\phi_h,\phi_S) &=& \frac{1}{P_T}\frac{Y_{\phi_h,\phi_S}-Y_{\phi_h,\phi_S +
    \pi}}{Y_{\phi_h,\phi_S} + Y_{\phi_h,\phi_S + \pi}} \nonumber \\
&\approx&  A_{C} \sin(\phi_h + \phi_S) + A_{S}\sin(\phi_h -
    \phi_S),  
\end{eqnarray}
 where $P_T$ is the target polarization, $\phi_h$ and $\phi_S$ are the
azimuthal angles of the hadron momentum and the target spin relative
to the lepton scattering plane (Fig.~\ref{fig:angle}), $Y$ is the normalized yield. 
$A_C$ is the Collins moment, which probes the convolution of the 
chiral-odd transversity distribution $h_1$ and the chiral-odd Collins fragmentation 
function (FF)\cite{collins}. $A_S$ is the Sivers moment, which probes the convolution of 
the naive T-odd quark Sivers function $f^{\perp}_{1T}$\cite{sivers1} and the unpolarized 
FF $D_1$\cite{Kotzinian:1995cz,tmd3}. While the transversity 
function describes the correlation of the quark spin with the nucleon spin (the probability 
of finding a transversely polarized parton inside a transversely polarized nucleon), 
the Sivers function reveals the correlation of the quark OAM with the nucleon spin. 
In particular, authors of Ref.~\refcite{siversOAM} provided constrains of quark angular momentum with world data of Sivers asymmetries and nucleon magnetic moments by assuming a connection between the generalized parton distribution E~\cite{GPDE} and the Sivers distribution.
Since
the Sivers function is odd under the naive time-reversal transformation,
it was originally believed to vanish\cite{collins3}. Recently, authors of Ref.~\refcite{brodsky}
showed that a nonzero $f_{1T}^{\perp}$ was possible due to QCD FSI between the struck 
quark and the residual nucleon system. It was further demonstrated that the Sivers
function that appears in the Drell-Yan process is the same as the one in SIDIS, but
with an opposite sign due to changes in the gauge link\cite{Collins_siver,brodsky_siver}. 
The experimental tests of above relation are crucial in order to demonstrate the validity 
of the QCD factorization theorem.

Another example is the angular dependence of the beam-helicity double-spin 
asymmetry (DSA) $A^{DSA}$ with a transversely polarized nucleon target and a longitudinal 
polarized lepton beam:
\begin{eqnarray}\label{eq:asy1}
A^{DSA}_{LT}\left(\phi_{h},\phi_{S}\right) & \equiv\frac{1}{P_{B}P_{T}}\frac{Y^{+}\left(\phi_{h},\phi_{S}\right)-Y^{-}\left(\phi_{h},\phi_{S}\right)}{Y^{+}\left(\phi_{h},\phi_{S}\right)+Y^{-}\left(\phi_{h},\phi_{S}\right)}\label{eq:ALT.Def}\\
 & \approx A_{LT}^{\cos\left(\phi_{h}-\phi_{S}\right)}\cos\left(\phi_{h}-\phi_{S}\right),\nonumber 
\end{eqnarray}
where $P_{B}$ is the polarization of the lepton beam, and $Y^{\pm}\left(\phi_{h},\phi_{S}\right)$
are the normalized yields for beam helicity states of $\pm1$. The first and
second subscripts to $A$ denote the respective polarization of beam
and target (L, T, and U represent longitudinal, transverse, and unpolarized,
respectively). Similar to the Collins and Sivers moments, the $A_{LT}^{\cos\left(\phi_{h}-\phi_{S}\right)}$
moment measures the convolution of $g_{1T}$ and the unpolarized FF $D_1$.
$g_{1T}$ has been calculated by lattice QCD\cite{lqcd_g1t,lqcd_g1t2} 
and in various quark models\cite{Jakob:1997wg,Pasquini:2008ax,Kotzinian:2008fe,Efremov:2009ze,Avakian:2010br,Bacchetta:2010si,Zhu:2011zz,Efremov:2011ye}.
In general, $g_{1T}^{u}$ ($g_{1T}^{d}$) is suggested to be positive (negative). 
In addition, $g_{1T}$ is shown to be closely connected with other TMDs through models. 
First, the $\boldsymbol{p}_{T}^{2}$-moment of $g_{1T}$ is linked to 
the collinear $g_1$ PDF through a Wandzura-Wilczek-type
approximation\cite{tmd1,Kotzinian:1995cz} in the leading twist. Second, 
$g_{1T}^{q}$ can be expressed as the combination of the quark transversity distribution $h_{1}^{q}$
and the pretzelosity distribution $h_{1T}^{\perp q}$ within the 
QCD parton model\cite{DiSalvo:2006wf}. Last, $g^q_{1T}$ is shown to be the same as $h_{1L}^{\perp q}$, but 
with an opposite sign in many models\cite{Lorce:2011zt} as a consequence of a geometric relation 
and some initial lattice calculations\cite{lqcd_g1t,lqcd_g1t2}.

On the experimental side, the HERMES and the COMPASS collaborations have performed pioneer 
work in measuring the SSA and the DSA. In particular, the HERMES collaboration carried out 
the first SSA measurement in SIDIS on a transversely polarized proton 
target using $e^\pm$ beams\cite{hermes-trans} at $Q^2=1.3-6.2$ GeV$^2$. 
The COMPASS collaboration performed the SIDIS measurements with a muon beam 
on transversely polarized deuteron\cite{compass-new} and 
proton\cite{compass-proton} targets at $Q^2=1.3-20.2$ GeV$^2$.
Both the HERMES\cite{hermes-new} and the COMPASS\cite{compass-proton}  proton data show significantly positive 
$\pi^+$ Sivers moments and close to zero $\pi^-$ Sivers moments, 
while results from the COMPASS deuteron data are consistent with zero. 
These measurements indicate a strong flavor dependence of the Sivers function\cite{ans_2012a}.
In addition, though the HERMES and the COMPASS
$\pi^+$ Sivers moments from their proton data 
are consistent in sign, potential discrepancies exist in magnitudes.
These discrepancies might be attributed to the energy evolution of the Sivers function\cite{sivers_evol,ans_2012a}.
For the Collins moment, large asymmetries were observed for both $\pi^+$ 
and $\pi^-$ from the proton data, but with opposite sign. These results show
that the ``unfavored'' Collins FF could be as large as the ``favored'' 
one\cite{collins3}, which is also consistent with the measured asymmetries of
the inclusive hadron pair production in the $e^+e^-$ annihilation from the
BELLE collaboration\cite{belle} (a direct measurement of product of Collins FFs). 
Furthermore, deuteron Collins asymmetries for $\pi^+$ and $\pi^-$ are
consistent with zero, which suggests a cancellation between proton and neutron. 
For the $A_{LT}$ DSA, the preliminary results of the COMPASS collaboration from
deuteron\cite{Parsamyan:2007ju} and proton\cite{Parsamyan:2010se} are 
consistent with zero within uncertainties.  The preliminary
results of proton from the HERMES collaboration\cite{hermes:alt} are also
mostly consistent with zero within uncertainties, except that a hint 
of positive $\pi^+$ moment is observed in the valence region. 

Due to the nature of the electromagnetic interaction and charges of the $u$ (2/3) and the $d$ (-1/3) quarks, 
the SIDIS measurement of a proton target would be dominated by the contribution
from $u$ quarks. To shed light on the flavor structure of TMDs,
it is important to perform the SSA and the DSA SIDIS measurements on
a neutron target, which is more sensitive to the $d$-quark contribution.
Since there is no stable free neutron target, polarized $^3$He is 
commonly used as an effective polarized neutron target\cite{he3_2}. 
The $^3$He nucleus is uniquely advantageous in extraction of information 
on the neutron spin compared to the deuteron (p+n), since
the ground state of $^3$He is dominated by the S state, where the two protons' 
spins cancel each other.


\section{Experiment E06-010}
The E06-010 experiment\cite{E06010} was carried out in Jefferson Lab (JLab) Hall A\cite{halla_nim} 
from 2008/11 to 2009/02. Measuring the SSA and the DSA in the pion electroproduction $n(e,e'\pi^{\pm})X$ 
on a transversely polarized $^3$He target is the main object of this experiment.  
The results of SSA and DSA have been reported in Ref.~\refcite{SSA_PRL} and Ref.~\refcite{DSA_PRL},
respectively. The beam energy of the longitudinally polarized electron was 5.89 GeV, the highest
available at that time. The average beam current throughout the entire experiment was 
12 $\mu A$ corresponding to a polarized luminosity of $\sim10^{36}N cm^{-2} s^{-1}$.
Polarized electrons in the beam were excited from a superlattice GaAs photocathode 
by a circularly polarized laser at the injector of the CEBAF accelerator. 
The electron  beam-helicity was flipped at 30 Hz by changing the laser polarization
with a Pockels cell. The beam polarization was periodically measured by a M$\o$ller polarimeter,
and the average value is determined to be $\left(76.8\pm3.5\right)\%$. 
For SSA measurements, unpolarized beam was achieved by summing the two beam helicity states, 
with the residual beam charge asymmetry smaller than 100 ppm per 1-hour run.

The 40 cm long $^3$He cell was filled with $\sim$8 atms of $^3$He 
and $\sim$0.13 atms of N$_2$ (reducing depolarization effects) at room temperature. 
The Spin Exchange Optical Pumping (SEOP) of a Rb-K mixture was used to polarize the 
$^3$He nuclei. The target polarization direction was controlled by 
three pairs of mutually orthogonal Helmholtz coils. Two orientations, 
vertical and horizontal polarizations in the plane transverse to
the beam direction was chosen to maximize the $\phi_S$ coverage.
The target spin was automatically flipped through the Adiabatic Fast 
Passage (AFP) every 20 minutes, while the target polarization was
measured by the Nuclear Magnetic Resonance (NMR) in each flip. 
The known water NMR signal was used to calibrate the NMR measurements, and
results were cross-checked with the Electron Paramagnetic
Resonance (SPR) method. The average polarization was $\left(55.4\pm2.8\right)\%$.

Scattered electrons with momenta from 0.6-2.5 GeV were detected in the BigBite 
spectrometer at a central angle of 30$^\circ$ on the beam right.
The BigBite spectrometer consisted of a large-opening dipole magnet  
in front of a detector stack including three sets of multi-wire drift chambers 
for tracking charged particles, a lead-glass calorimeter divided 
into preshower/shower sections for identifying electrons, and a 
scintillator plane between the preshower and shower for determining timing. 
The average solid angle was $\sim$64 msr. The large out-of-plane angle
acceptance of the BigBite ($\pm$240 mrad) spectrometer 
was essential in maximizing the $\phi_h$ coverage of the experiment. 
The optics property of the BigBite magnet was 
calibrated using a multi-foil carbon target, a sieve slit collimator and the 
$^1$H$(e,e')p$ elastic scattering at incident energies of 1.2 and 2.4 GeV. 
The achieved angular and momentum resolutions were better than 10 mrad and 1\%, 
respectively. A clean $e^-$ identification was achieved using cuts on the preshower 
energy $E_{ps}$ and the ratio $E/p$ of the total shower energy to the momentum from optics 
reconstruction.

Coincident charged hadrons were detected in the High Resolution Spectrometer (HRS)
at a central angle of 16$^\circ$ on the left side of beam 
and a central momentum of 2.35 GeV. 
The HRS detector package was configured to detect hadrons\cite{halla_nim}. 
The pion identification was achieved by combining a light gas \v{C}erenkov, 
a lead glass calorimeter, an aerogel \v{C}erenkov detector, and 
the time-of-flight (TOF) information.

SIDIS events were selected using cuts on the four-momentum
transfer squared $Q^2 > $ 1 GeV$^2$, the hadronic final-state
invariant mass W $>$ 2.3 GeV, and the mass of undetected
final-state particles W$^{\prime}$ $>$ 1.6 GeV, 
assuming that the electron is scattered on a nucleon. The raw Collins and Sivers moments
were obtained by fitting the asymmetries in 2-D ($\phi_h,
\phi_S$) bins according to Eqn.\eqref{eq:asy}. The raw $A_{LT}$ moment was obtained
by an unbinned maximum-likelihood method according to Eqn.~\eqref{eq:asy1}.
Results were further cross-checked with both methods. 
The neutron SSA and DSA were extracted from the measured $^3$He moments after 
correcting the directly measured N$_2$ dilution through
\begin{equation}
A_{\rm ^3He}^{SSA(DSA)} = P_n \cdot (1-f_p) \cdot A_{n}^{SSA(DSA)}  + P_p f_p
\cdot A_{p}^{SSA(DSA)} \label{eqn:effpol}.
\end{equation}
Here,  $P_n=0.86^{+0.036}_{-0.02}$ ($P_p=-0.028^{+0.009}_{-0.004}$) is the neutron
(proton) effective polarization\cite{Zheng_PRC}. The $f_{p}= \frac{2\sigma_p}{\sigma_{\rm ^3He}}$ 
is referred to as the proton dilution which is directly measured
by comparing the yields of unpolarized hydrogen and $^3$He targets. 
This approach was validated by a theoretical calculation\cite{Scopetta}. 

\begin{figure}[]
\centering
\includegraphics[width=120mm]{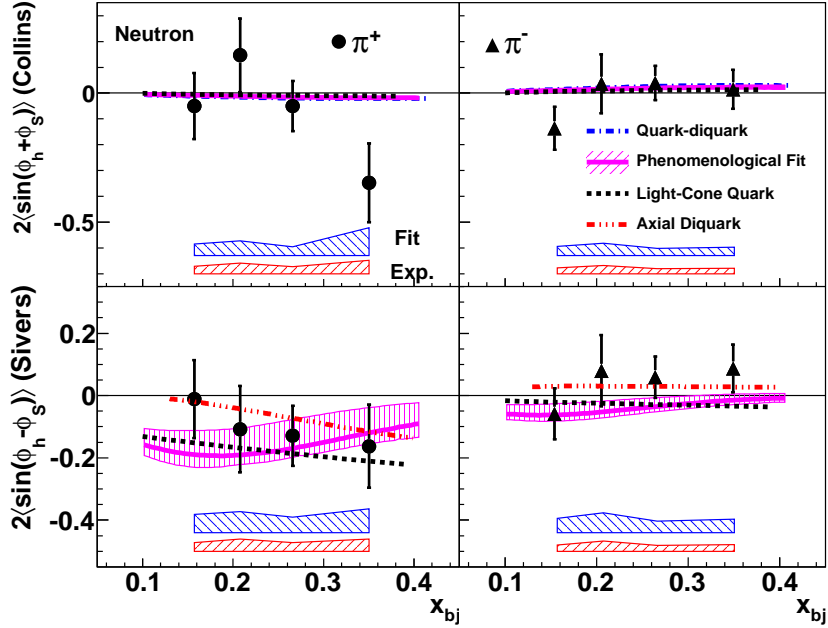}
\caption{\label{fig:neutron} (Color online) The extracted neutron Collins and 
 Sivers moments with uncertainty bands for both $\pi^+$ and $\pi^-$
 electro-production from Ref.~\protect\refcite{SSA_PRL}. See text for details.}  
\end{figure}

The background in the SIDIS electron sample that comes from $e^+/e^-$ pair production
is the largest systematic uncertainty. This background was directly measured 
by reversing the polarity of the BigBite magnet to detect $e^+$ in 
identical conditions as $e^-$. Other experimental 
uncertainties include the $K^\pm$ contamination in the $\pi^\pm$ sample, 
effects of bin-centering/resolution/radiative corrections, the effect of the 
target collimator, the fluctuation of the target density, the false asymmetry
due to the radiation damage to the BigBite preshower calorimeter,
the contamination from the diffractive $\rho$ meson production, 
the contamination from radiative tails of the exclusive electroproduction,
and the effect of the $\pi^{\pm}$ final state interaction. The quadrature sum of all 
above contributions is below 25\% of the statistical uncertainty.

The extracted neutron Collins and Sivers moments\cite{SSA_PRL} are shown in the
Fig.~\ref{fig:neutron}. The fitting systematic uncertainty (red band) is 
due to the neglect of other $\phi_h$- and $\phi_S$-dependent terms, 
such as the $2\langle\sin (3\phi_h-\phi_S)\rangle$, 
higher-twist terms including the $2\langle\sin \phi_S\rangle$ and the
$2\langle\sin\left(2\phi_h-\phi_S\right)\rangle$, azimuthal
modulations of the unpolarized cross section including the 
Cahn ($2\langle\cos \phi_h\rangle$) and the Boer-Mulders (2$\langle\cos (2\phi_h)\rangle$) 
terms. Collins moments are compared with the phenomenological fit\cite{Anselmino:2007fs}, 
a light-cone quark model calculation\cite{LCQM_C1,LCQM_C2} and quark-diquark model
\cite{CQM_C1,CQM_C2} calculations. With the Soffer's bound\cite{soffer}, the 
phenomenological fit and model calculations predict rather small 
asymmetries which mostly agree with data, except the 
$\pi^+$ Collins moment at $x=0.34$. Negative $\pi^+$ Sivers moments 
are favored by these data, while the $\pi^-$ moments are close to zero.
Within the parton model picture, such behavior is consistent with 
a negative $d$ quark Sivers function, which has been suggested by 
predictions of the phenomenological fit\cite{anselmino3}
to the HERMES and the COMPASS data, a light-cone quark model calculation\cite{LCQM_S1,LCQM_S2},
and an axial diquark model calculation\cite{ADQ}.

\begin{figure}[]
\centering
\includegraphics[width=120mm]{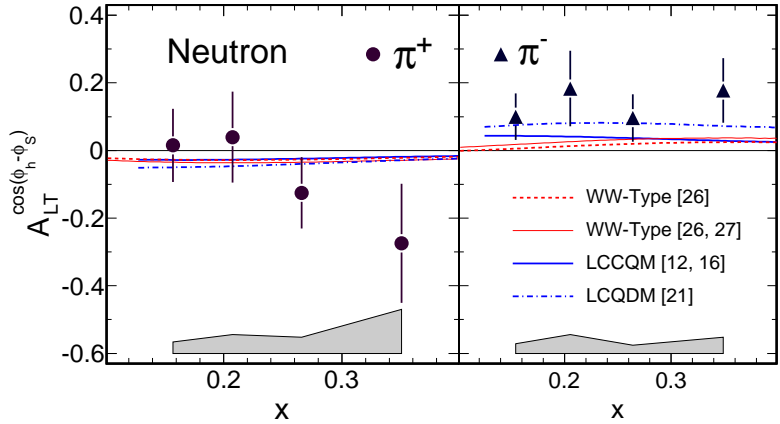}
\caption{\label{fig:neutron_ALT} (Color online) The neutron $A_{LT}^{\cos\left(\phi_{h}-\phi_{S}\right)}$
azimuthal asymmetry for positive (left) and negative (right) charged
pions vs $x$ from Ref.~\protect\refcite{DSA_PRL}. }  
\end{figure}

The extracted neutron $A_{LT}$ moments are shown in the Fig.~\ref{fig:neutron_ALT}. 
Beside aforementioned experimental systematic uncertainties, additional 
uncertainties include: the contamination from the DSA $A_{LL}$ due to a 5\%-20\% 
longitudinal component of the target polarization with respect to the virtual-photon
direction and the contribution from the $A^p_{LT}$ estimated from the COMPASS 
preliminary data\cite{Parsamyan:2010se}. Several model calculations, 
including WW-type approximations with parametrizations from Ref.~\refcite{Kotzinian:2006dw} and 
Ref.~\refcite{Prokudin.DraftPaper}, a light-cone
constituent quark model (LCCQM)\cite{Pasquini:2008ax,Boffi:2009sh}, 
and a light-cone quark-diquark model (LCDQM) evaluated
using the second approach in the Ref.~\refcite{Zhu:2011zz}, are plotted as well. 
While the extracted $A_{LT}^{n}\left(\pi^{+}\right)$ is consistent with zero within
uncertainties, the results of $A_{LT}^{n}\left(\pi^{-}\right)$ are consistent in
sign with these model predictions but favor a larger magnitude. 
The sign of $A_{LT}^{n}\left(\pi^{-}\right)$ is opposite to the sign
of the $A_{UL}^{\sin2\phi_{h}}$ asymmetry in the $\pi^{+}$ production
on the proton measured by the CLAS collaboration\cite{Avakian:2010ae}, which is 
predicted by many models\cite{Lorce:2011zt}.

\section{Future Opportunities}
The current generation of experiments including the HERMES, the COMPASS, and the E06-010 
play important roles in exploring TMDs. However, compared to the collinear unpolarized and longitudinal
polarized functions ($f_1$ and $g_1$) which depend on $x$ and $Q^2$ only, TMDs are much less 
understood due to their multi-dimensional nature ($x$, $Q^2$ and the quark transverse momentum $p_T$).
For example, the kinematics of $x$, $z$ and $P_T$ are always
strongly correlated in all existing experiments, and results are
usually shown in one dimensional format ($x$, $z$ or $P_T$) with
integration over the other two variables. Furthermore, many assumptions, such as a Gaussian
approximation of $p_T$ dependence, have been adopted in the global fit to limit the total
number of parameters.  Therefore, in order to improve our understanding on TMDs and to
reduce aforementioned theoretical assumptions, it is important to perform
precision measurements in multiple dimensions. 

\begin{figure}[]
\centering
\includegraphics[width=100mm]{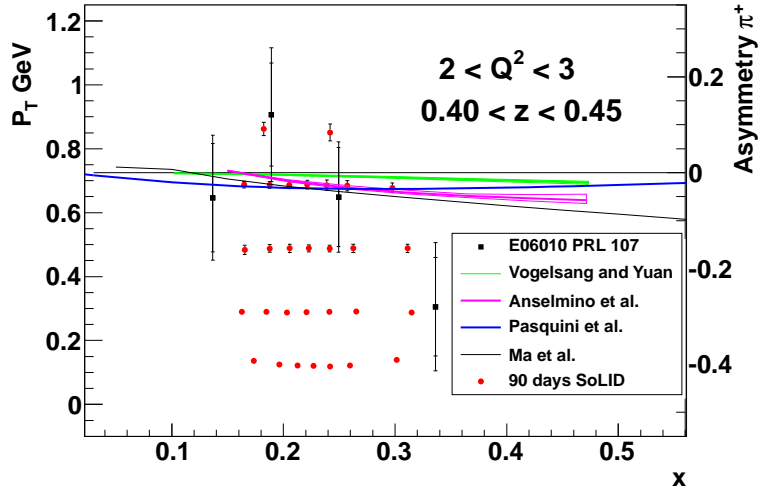}
\caption{\label{fig:12gev} (Color online) 12 GeV projections with the SoLID spectrometer 
and the polarized $^3$He target of $\pi^+$ Collins asymmetries 
at $0.4<z<0.45$ and 2 GeV$^2<Q^2>3$ GeV$^2$. The Position of 
each projected point on the left y-axis represents 
the average $P_T$ value of the corresponding kinematic bin.  
The statistical uncertainty of each point and magnitudes of
theoretical calculations follow the right y-axis.}  
\end{figure}

Jefferson Lab is upgrading its incident electron beam energy to 12 GeV, which
provides unique opportunities to carry out measurements of semi-inclusive 
hadron yields from deep-inelastic scattering. Two experiments\cite{SoLID_nT,SoLID_nL} 
have been approved with the same polarized $^3$He target and a large 
acceptance solenoid device (SoLID)\cite{trans_12gev}. The fixed-target
kinematics allows for a probe of the interesting high $x$ region (0.05-0.65), 
which is essential in determining quark tensor charges. In addition, the 
full azimuthal angular coverage of SoLID results in a significant reduction
of systematic uncertainties of the luminosity, detection efficiencies, etc.,
which is essential for high precision measurements. The projected 
results for $\pi^+$ Collins asymmetry are shown in the
Fig.~\ref{fig:12gev} for one typical kinematic bin (48 bins in total), 
0.4 $<$ $z$ $<$ 0.45, 2 GeV$^2$ $<$ $Q^2$ $<$ 3 GeV$^2$. 
Theoretical predictions of the Collins asymmetries from Anselmino {\it et al.}\cite{ansel_priv}, 
Vogelsang and Yuan~\cite{yuan_priv}, Ma {\it et al.}\cite{Huang}, 
and Psaquini {\it et al.}\cite{Pasquini:2008ax} are shown as well. 
The results of E06-010~\cite{SSA_PRL} are shown as black points.
In addition, an experiment with a 
transversely polarized proton target\cite{SoLID_p} is proposed  to achieve the flavor separation. These 
next generation experiments will provide a high precision measurement 
of various SSA and DSA asymmetries in the valence quark region.

\begin{figure}[]
\centering
\includegraphics[width=100mm]{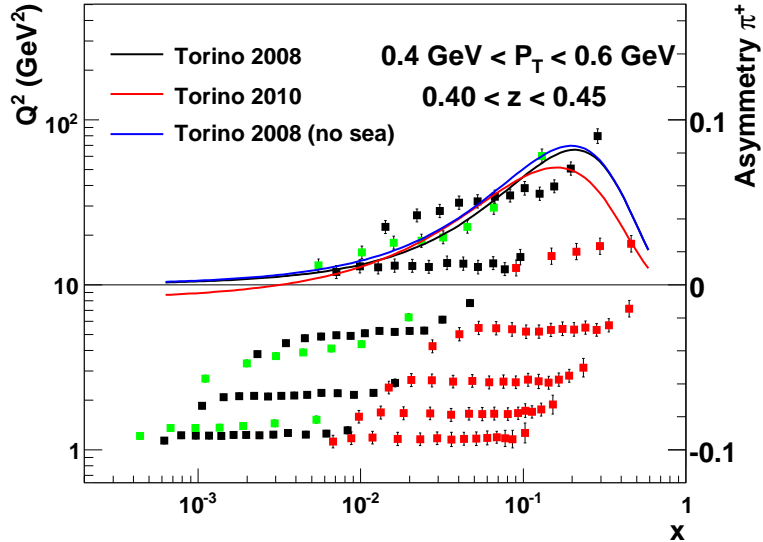}
\caption{\label{fig:eic} (Color online) Projections of the Sivers asymmetry of the 
$\pi^+$ electro-production off the proton\protect\cite{trans_EIC} in 
a P$_T$ and $z$ bin out of 40 bins along with various calculated asymmetries. 
The black, green, and red dots represent the 11+60 GeV, the 11+100 GeV, and the 3+20 GeV 
EIC configurations, respectively. The integrated luminosity is assumed to be 3.1$\times$10$^{40}$ $cm^{-2}$,
9.3$\times$10$^{40}$ cm$^{-2}$, and 3.1$\times$10$^{40}$ cm$^{-2}$ for 3+20 GeV, 11+60 GeV
and 11+100 GeV configurations, respectively. The $x$-axis represents Bjorken $x$, and 
the left $y$-axis represents $Q^2$. The position of each point represents the 
position of the kinematic bin in the $x$-$Q^2$ phase space. The error bar of each 
point and magnitudes of calculated asymmetries follow the right $y$-axis. }  
\end{figure}
\begin{figure}[]
\centering
\includegraphics[width=120mm]{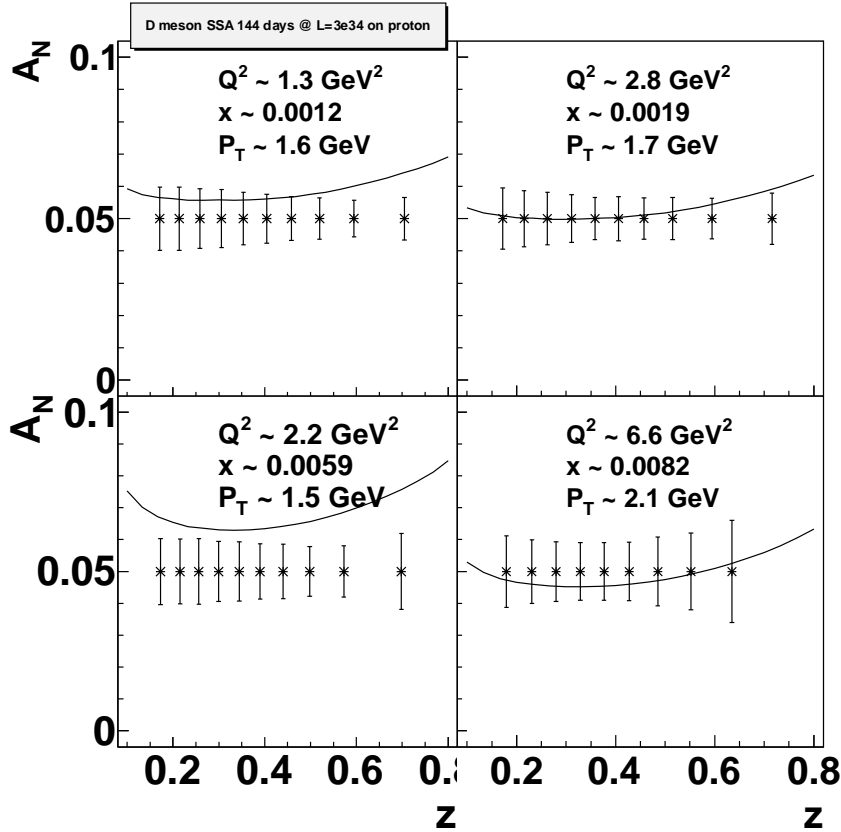}
\caption{\label{fig:eic_dmeson} The projected results\protect\cite{eic} on the transverse SSA 
of the $D$ meson production. The data are binned 2-by-2 in terms of $x$ and $Q^2$. 
Within each $x$-$Q^2$ bin, projections are plotted with $z$. 
The central kinematics are listed in each panel.}  
\end{figure}

Compared to the fixed target experiments, an electro-ion collider (EIC) will be 
able to probe much larger phase space in $x$, $Q^2$ and $P_T$ due to a much large center 
of mass energy. In particular, an EIC is an ideal machine to study TMDs for 
sea quarks and gluons at low $x$. The large $Q^2$ coverage will 
allow a study of the evolution of TMDs as well as various higher twist effects. 
In addition, the large hadron transverse momentum $P_T$ coverage will allow a study of 
SSA and DSA phenomenon at high $P_T$, where NLO QCD processes
dominate. Furthermore, an EIC enables the capability to detect open charm mesons. 
These new probes open a window to study gluon distribution functions~\cite{trans_EIC}. For example, 
SSAs of the open-flavor (anti)D meson production in the DIS regime provide a unique 
opportunity to measure tri-gluon correlation functions~\cite{Kang:2008qh,Beppu:2010qn}, which 
are closely connected to the gluon's transverse motion and the color coherence inside a 
transversely polarized nucleon. Summaries of the physics potential of an EIC are 
presented in the Ref.~\refcite{trans_EIC} and the Ref.~\refcite{eic}.  
Fig.~\ref{fig:eic} shows the expected projection of the $\pi^+$ Sivers 
asymmetry with a proton beam at a high luminosity EIC in the kinematics bin of $0.4<z<0.45$
 and $0.4~ {\rm GeV}<P_T<0.6~{\rm GeV}$. Together with the projection, 
a few calculations of the Sivers asymmetry from Ref.~\refcite{ansel_priv} and 
Ref.~\refcite{Anselmino:2008sga} are presented.  Fig.~\ref{fig:eic_dmeson} shows the projection for 
the $D$ transverse SSA measurement for a running time of 144 days with a proton beam 
at a luminosity of $3\times 10^{34}$cm$^{-2}$s$^{-1}$ together with a theoretical prediction from 
the Ref.~\cite{Kang:2008qh}. With the 12 GeV upgrade and a high luminosity electron-ion 
collider, the knowledge of TMDs of gluons and quarks can be greatly advanced, 
and would ultimately improve our understanding of TMDs from the first principle of QCD. 

\section{Summary}
The Jefferson Lab E06-010 experiment measured the neutron SSA and DSA for the first time 
through pion electroproduction in the DIS region with a transversely polarized $^3$He
target. These data provide valuable information about nucleon TMDs, which 
describes not only nucleon structure in 3-D momentum space, but also provide insights to 
the dynamics of QCD in the confinement region. The precision measurements at 12-GeV 
Jefferson Lab and a future EIC would ultimately realize the multi-dimensional
mapping of TMDs, which will bring our current understanding of the nucleon 
structure to a completely new level.

\section*{Acknowledgments}
This work was supported in part by Caltech, U.S. Department of 
Energy (DOE) under contract number DE-FG02-03ER41231, the National Science 
Foundation, and by DOE contract number DE-AC05-06OR23177, under which
the Jefferson Science Associates (JSA) operates the Thomas Jefferson 
National Accelerator Facility.


\begin{thebibliography}{0}
\bibitem{asy1} D. J. Gross and F. Wilczek, Phys. Rev. Lett. {\bf 30}, 1343 (1973).
\bibitem{asy2} H. D. Politzer, Phys. Rev. Lett. {\bf 30}, 1346 (1973).
\bibitem{hera1} F. Caola, S. Forte, and J. Rojo, Nucl. Phys. {\bf A854}, 32 (2011).
\bibitem{spin} S. E. Kuhn, J.-P. Chen, and E. Leader, Prog. Part. Nucl. Phys. {\bf 63}, 1 (2009). 
\bibitem{trans1} K. Hidaka, E. Monsay and D. Sivers, Phys. Rev. {\bf D19}, 1503 (1979).
\bibitem{trans2} J. P. Ralston and D. E. Soper, Nucl. Phys. {\bf B152}, 109 (1979).
\bibitem{trans3} R. L. Jaffe and X. Ji, Phys. Rev. Lett. {\bf 67}, 552 (1991).
\bibitem{tc_lqcd} M. Gockeler {\it et al.}, Phys. Lett. {\bf B627}, 113 (2005).
\bibitem{tc_mod1} H.-X. He and X. Ji, Phys. Rev. {\bf D52}, 2960 (1995).
\bibitem{tc_mod2} V. Barone {\t et al.}, Phys. Lett. {\bf B390}, 287 (1997).
\bibitem{tc_mod3} B.-Q. Ma and I. Schmidt, J. Phys. {\bf G24}, 71, (1998).
\bibitem{tc_mod4} B. Pasquini {\it et al.}, Phys. Rev. {\bf D72}, 094029 (2005).
\bibitem{tc_mod5} M. Wakamatsu, Phys. Lett. {\bf B653}. 398 (2007).
\bibitem{tc_mod6} I. C. Cloet {\it et al.}, Phys. Lett. {\bf B659}, 214 (2008).
\bibitem{soffer} J. Soffer, Phys. Rev. Lett. {\bf 74}, 1292 (1995).
\bibitem{Kumano:1997qp} S. Kumano and M. Miyama, Phys. Rev. {\bf D56}, 2504 (1997).
\bibitem{Hayashigaki:1997dn} A. Hayashigaki {\it et al.}, Phys. Rev. {\bf D56}, 7350 (1997).
\bibitem{werner} W. Vogelsang, Phys. Rev. {\bf D57}, 1886 (1998).
\bibitem{doubt} J. P. Ralston, arXiv:0810.0871 (2008).
\bibitem{tmd1} P. J. Mulders and R. D. Tangerman, Nucl. Phys. {\bf B461}, 197 (1996).
\bibitem{tmd2} D. Boer, P. J. Mulders, Phys. Rev. {\bf D57}, 5780 (1998).
\bibitem{tmd3} A. Bacchetta {\it et al.}, JHEP {\bf 02}, 093 (2007).
\bibitem{largenc} P. V. Pobylitsa, hep-ph/0301236.
\bibitem{brodsky} S. J. Brodsky, D. S. Hwang and I. Schmidt, Phys. Lett. {\bf B530}, 99 (2002).
\bibitem{oam} X. Ji, J.-P. Ma, and F. Yuan, Nucl. Phys. {\bf B652}, 383 (2003).
\bibitem{trento} A. Bacchetta {\it et al.}, Phys. Rev. {\bf D70}, 117504 (2004).
\bibitem{collins} J. C. Collins, Nucl. Phys. {\bf B396}, 161 (1993).
\bibitem{sivers1} D. Sivers, Phys. Rev. {\bf D41}, 83 (1990).
\bibitem{Kotzinian:1995cz} A. M. Kotzinian and P. J. Mulders, Phys. Rev. {\bf D54}, 1229 (1996).
\bibitem{siversOAM} A. Bacchetta and M. Radici, arXiv:1107.5755.
\bibitem{GPDE} X. Ji, Phys. Rev. Lett. {\bf 78}, 610 (1997).
\bibitem{collins3} J. C. Collins, Nucl. Phys. {\bf B396}, 161 (1993).
\bibitem{Collins_siver} J. C. Collins, Phys. Lett. {\bf B536}, 43 (2002).
\bibitem{brodsky_siver} S. J. Brodsky {\it et al.}, Nucl. Phys. {\bf B642}, 344 (2002).
\bibitem{lqcd_g1t} P. Hagler {\it et al.}, Europhys. Lett. {\bf 88}, 61001 (2009).
\bibitem{lqcd_g1t2} B. U. Musch {\it et al.}, Phys. Rev. {\bf D83}, 094507 (2011).
\bibitem{Jakob:1997wg} R. Jakob, P. Mulders, and J. Rodrigues, Nucl. Phys. {\bf A626}, 937 (1997).
\bibitem{Pasquini:2008ax} B. Pasquini, S. Cazzaniga, and S. Boffi, Phys. Rev. {\bf D78}, 034025 (2008).
\bibitem{Kotzinian:2008fe} A. Kotzinian, arXiv:0806.3804 (2008).
\bibitem{Efremov:2009ze} A. V. Efremov {\it et al.}, Phys. Rev. {\bf D80}, 014021 (2009).
\bibitem{Avakian:2010br} H. Avakian {\it et al.}, Phys. Rev. {\bf D81}, 074035 (2010).
\bibitem{Bacchetta:2010si} A. Bacchetta {\it et al.}, Euro. Phys. J. {\bf A45}, 373 (2010).
\bibitem{Zhu:2011zz} J. Zhu and B.-Q. Ma, Phys. Lett. {\bf B696}, 246 (2011).
\bibitem{Efremov:2011ye} A. V. Efremov {\it et al.}, J. Phys. Conf. Ser. {\bf 295}, 012052 (2011).
\bibitem{DiSalvo:2006wf} E. Di Salvo, Mod. Phys. Lett. {\bf A22}, 1787 (2007).
\bibitem{Lorce:2011zt} C. Lorce and B. Pasquini, Phys. Rev. {\bf D84}, 034039 (2011).
\bibitem{hermes-trans} A. Airapetian {\it et al.}, Phys. Rev. Lett. {\bf 94}, 012002 (2005).
\bibitem{compass-new} M. Alekseev {\it et al.}, Phys. Lett. {\bf B673}, 127 (2009).
\bibitem{compass-proton} M. Alekseev {\it et al.}, Phys. Lett. {\bf B692}, 240 (2010).
\bibitem{hermes-new}A. Airapetian {\it et al.}, Phys. Rev. Lett. {\bf 103}, 152002 (2009).
\bibitem{ans_2012a} M. Anselmino, M. Boglione, and S. Melis, arXiv:1204.1239 (2012).
\bibitem{sivers_evol} S. M. Aybat {\it et al.}, Phys. Rev. {\bf D85},034043 (2012).
\bibitem{belle} R. Seidl {\it et al.} Phys. Rev. Lett. {\bf 96}, 232002 (2006).
\bibitem{Parsamyan:2007ju}  B. Parsamyan (COMPASS), Euro. Phys. J. Special Topics {\bf 162}, 89 (2008).
\bibitem{Parsamyan:2010se} B. Parsamyan (COMPASS), J. Phys. Conf. Ser. {\bf 295}, 012046 (2011).
\bibitem{hermes:alt} L. L. Pappalardo and M. Diefenthaler (HERMESS), arXiv:1107.4227 (2011).
\bibitem{he3_2} F. Bissey {\it et al.} Phys. Rev. {\bf C65}, 064317 (2002).
\bibitem{E06010} Jefferson Lab E06010, Spokesperson: J.-P. Chen, E. Cisbani, H. Gao, X. Jiang, and J.-C. Peng.
\bibitem{halla_nim} J. Alcorn {\it et al.}, Nucl. Instr. and Meth. {\bf A522}, 294 (2004).
\bibitem{SSA_PRL} X. Qian {\it et al.} Phys. Rev. Lett. {\bf 107}, 072003 (2011).
\bibitem{DSA_PRL} J. Huang {\it et al.} Phys. Rev. Lett. {\bf 108}, 052001 (2012).
\bibitem{Zheng_PRC} X. Zheng {\it et al.}, Phys. Rev. {\bf C70}, 065207 (2004).
\bibitem{Scopetta} S. Scopetta, Phys. Rev. {\bf D75}, 054005 (2007).
\bibitem{Anselmino:2007fs} M. Anselmino {\it et al.}, Phys. Rev. {\bf D75}, 054032 (2007).
\bibitem{LCQM_C1} S. Boffi {\it et al.}, Phys. Rev. {\bf D79}, 094012 (2009).
\bibitem{LCQM_C2} B. Pasquini, S. Cazzaniga and S. Boff, Phys. Rev. {\bf D78}, 034025 (2008).
\bibitem{CQM_C1} J. She and B. Q. Ma, Phys. Rev. {\bf D83}, 037502 (2011).
\bibitem{CQM_C2} B. Q. Ma, I. Schmidt, J. J. Yang, Phys. Rev. {\bf D65}, 034010 (2002). 
\bibitem{anselmino3} M. Anselmino {\it et al.}, Phys. Rev. {\bf D72}, 094007 (2005).
\bibitem{LCQM_S1} S. Arnold {\it et al.}, arXiv:0805.2137 (2008).
\bibitem{LCQM_S2} B. Pasquini and P. Schweitzer, Phys. Rev. {\bf D83}, 114044 (2011).
\bibitem{ADQ} L. P. Gamerg {\it et al.} Phys. Rev. {\bf D77}, 094016 (2008).

\bibitem{Kotzinian:2006dw} A. Kotzinian, B. Parsamyan, and A. Prokudin, Phys. Rev. {\bf D73}, 114017 (2006)
\bibitem{Prokudin.DraftPaper} A. Prokudin {\it private communications}.
\bibitem{Boffi:2009sh} S. Boffi {\it et al.}, Phys. Rev. {\bf D79}, 094012 (2009).
\bibitem{Avakian:2010ae} H. Avakian {\it et al.} (CLAS), Phys. Rev. Lett. {\bf 105}, 262002 (2010).
\bibitem{SoLID_nT} Jefferson Lab E12-10-006, Spokesperson: J.-P. Chen, H. Gao, X. Jiang, X. Qian, J.-C. Peng.
\bibitem{SoLID_nL} Jefferson Lab E12-11-007, Spokesperson: J.-P. Chen, J. Huang, Y. Qiang, W. B. Yan.
\bibitem{trans_12gev} H. Gao {\it et al.} Enr. Phys. J. Plus {\bf 126}, 2 (2011).
\bibitem{ansel_priv} M. Anselmino and A. Prokudin, {\it private communications}.
\bibitem{yuan_priv} W. Vogelsang and F. Yuan, {\it private communications}.
\bibitem{Huang}Y. Huang, J. She and B.-Q. Ma, Phys. Rev. {\bf D76}, 034004 (2007).
\bibitem{SoLID_p} Jefferson Lab PR12-11-108, Spokesperson: K. Allada, J.-P. Chen, X. M. Li, H. Gao, Z.-E. Meziani. 

\bibitem{trans_EIC} M. Alselmino {\it et al.},  Euro. Phys. J. {\bf A47}, 35 (2011).
\bibitem{Kang:2008qh} Z. B. Kang and J. W. Qiu, Phys. Rev. {\bf D78}, 034005 (2008).
\bibitem{Beppu:2010qn} H. Beppu {\it et al}, Phys. Rev. {\bf D82}, 054005 (2010).
\bibitem{eic} D. Boer {\it et al.} arXiv:1108.1713 (2011).
\bibitem{Anselmino:2008sga} M. Anselmino {\it et al.}, Euro. Phys. J. {\bf A39}, 89 (2009).



\end{thebibliography}
\end{document}